\documentclass[12pt]{article}
\usepackage{graphicx}
\usepackage{amssymb}
\usepackage{latexsym}
\usepackage{amsmath}
\usepackage{epstopdf}
\def\thefootnote {\alph{footnote})}

\newcommand{\beq}{\begin{equation}}
\newcommand{\eeq}{\end{equation}}

\font\mybbi=msbm10 at 9pt
\def\bbi#1{\hbox{\mybbi#1}}

\def\R{\mathbb{R}}
\def\N{\mathbb{N}}
\def\C{\mathbb{C}}

\def\NU{\Upsilon}

\def\_\BC{\bbi C}

\def\SD{\rtimes}

\def\dS {de Sitter}
\def\ds {de Sitter}
\def\R{{\rm I\!R}}

\newcommand{\vka}{\varkappa}
\newcommand{\vth}{\vartheta}
\newcommand{\vsi}{\varsigma}

\newcommand{\be}{\beta}
\newcommand{\ga}{\gamma}

\newcommand{\al}{\alpha}

\newcommand{\lga}{\longrightarrow}

\newcommand{\bra}{\begin{array}}
\newcommand{\era}{\end{array}}
\newcommand{\bqn}{\begin{eqnarray}}
\newcommand{\eqn}{\end{eqnarray}}
\newcommand\ben{\begin{enumerate}}
\newcommand\een{\end{enumerate}}
\newcommand\bei{\begin{itemize}}
\newcommand\ei{\end{itemize}}

\title{The nature of $\Lambda$  and the mass of the graviton:\\
 A critical view}
\author{J. P. Gazeau$^{\ast}$ and M. Novello$^{\ast\ast}$ \\
$^{\ast}$ Laboratoire
Astroparticules et Cosmologie (APC, UMR 7164),\\ Boite 7020
Universit\'e Paris 7 Denis Diderot,\\ 2 Place Jussieu, 75251 Paris
Cedex 05 Fr\\
{\small  gazeau@ccr.jussieu.fr}\\
$^{\ast\ast}$ ICRA, Centro Brasileiro de Pesquisas Fisicas,\\
Rua Dr. Xavier Sigaud 150,\\
CEP 22290-180, Rio de Janeiro, Brazil\\
{\small  novello@cbpf.br}}
\date{\today}
\begin{document}
 \maketitle

\begin{abstract}
The existence of a non-zero cosmological constant $\Lambda$
gives rise to controversial interpretations. Is $\Lambda$ a
universal constant fixing the geometry of an empty universe, as fundamental as the Planck constant or the speed of light in the vacuum? Its natural place is then on the
left-hand side of the Einstein equation.  Is it instead something emerging
from a perturbative calculus performed on the metric $g_{\mu\nu}$ solution
of the Einstein equation and to which it might be given a material status
of (dark or bright) ``energy"?  It should then be part of the content of the right-hand
side of the Einstein equations.  The purpose of this paper is not to elucidate the fundamental nature of $\Lambda$, but instead we aim  to present and
discuss some of the arguments in favor of both interpretations of the cosmological constant.   We will analyse the
question of a $\Lambda$-dependent graviton mass, more precisely  the possibility
that between the Compton wavelength of the graviton and the
cosmological constant there is the relation $l_{g} \,
\Lambda^{\frac{1}{2}} \approx 1.$ 
Since a physical quantity like  mass originates in a minkowskian conservation law, we  proceed to a group theoretical
interpretation of this relation in terms of the two possible
$\Lambda$-deformations of the Poincar\'e group, namely the de Sitter
and anti de Sitter groups. We  use a very suitable  formula, the so-called Garidi mass, and the typically dS/AdS dimensionless parameter $\hbar\, H/ mc^2$  in order to make clear the  asymptotic relations between minkowskian masses $m$ and their possible dS/AdS counterparts. We conclude that if the fundamental of the geometry of space-time is minkowskian, then the square of the mass of the graviton is proportional to $\Lambda$; otherwise, if the fundamental state is deSitter/AdS, then the graviton is massless in the deSitterian sense. 

\end{abstract}


\renewcommand{\thefootnote}{\arabic{footnote}}
\section{Introduction}

\subsection{Two Einstein theories of gravity}

Let us make a \emph{Gedankenexperiment} to examine what should be the changes
concerning the interpretation of the gravitational field if the
cosmological constant were to be viewed as a fundamental quantity
that is not related to any property, classical or quantum, of the
matter. It is understood that we are dealing here with some sort
of bare cosmological constant, and not with the observed one that
should contain modifications coming from the classical or the
quantum fluctuations of matter fields. From the very beginning one
should clearly state that there are two distinct theories proposed
by Einstein to represent the dynamics of the geometry, theories that we
name EGR-1 and EGR-2 and which were elaborated by Einstein to deal  respectively
with local gravitational phenomena and within a cosmological
context. One can distinguish between these theories as follows:

\paragraph{Theory {\rm EGR-1}}
\begin{equation}
R_{\mu\nu} - \frac{1}{2} \, R \, g_{\mu\nu} = -\kappa \,
T_{\mu\nu}. \label{27nov3}
\end{equation}

From this point of view, the corresponding fundamental state that contains
the maximum number of symmetries is the minkowskian geometry.

\paragraph{Theory {\rm EGR-2}}
\begin{equation}
R_{\mu\nu} - \frac{1}{2} \, R \, g_{\mu\nu} + \Lambda \,
g_{\mu\nu} = -\kappa \, T_{\mu\nu}. \label{27nov3}
\end{equation}

In EGR-2 theory, the corresponding fundamental state that contains
the maximum number of symmetries is the de-Sitter/anti-de-Sitter
geometry.

On certain aspects, such a fundamental $\Lambda$
 can be considered as intimately
related to what one should call the mass of the graviton
\cite{Novello-Neves}. The idea behind such a link between the
cosmological constant and the mass of the graviton should be more
clearly understood if we analyze  the structure of the kinematical
groups underlying these two theories, that is the
Poincar\'e-Lorentz and the de Sitter groups, respectively. We
shall see that the concept of mass, that is associated to one of
the two Casimir invariants of the Poincar\'e-Lorentz group does
not have, in general, a counterpart in the de Sitter groups. Let
us here recall the basic features of the latter.

\subsection{De Sitter and anti de Sitter space-times and groups}

De Sitter and anti de Sitter space-times play a fundamental role in
cosmology, since they are, with Minkowski, the only maximally symmetric
space-time solutions in general relativity. Their
respective invariance (in the relativity or kinematical sense) groups
are the ten-parameter \ds ~$SO_{\scriptsize 0}(1,4)$ and anti \ds ~$SO_{\scriptsize 0}(2,3)$ groups.
Both may be seen as   deformations of the
proper orthochronous Poincar\'e group $\R^{1,3}\SD\, S0_{\scriptsize
0}(1,3)$, the kinematical group of Minkowski.

As recalled above, the de Sitter [resp. anti de Sitter]  space-times are  the  unique
maximally symmetric solutions of the vacuum Einstein's equations with positive [resp.
negative] cosmological constant $\Lambda$. This constant is linked
to the (constant) Ricci  curvature $4 \Lambda$ of these
space-times and it allows to introduce the fundamental curvature
or inverse length $\vka = H c=\sqrt{\vert \Lambda\vert/3}$, ($H$ is
the Hubble constant).

These   space-times may be viewed  as forming a one-parameter family
of  deformations of the Minkowski space-time.

Serious reasons back up any interest in studying Physics in
such constant curvature spacetimes with maximal symmetry. The first
one is the simplicity of their geometry, which makes consider them as
an excellent laboratory model in view of studying Physics in more
elaborate universes, more precisely with the purpose to set up a quantum field theory as much rigorous as possible \cite{ISH,[BD],FUL,WALD}.

Higher dimensional anti de Sitter spaces have becoming in the last
years very popular because of their regularizing geometries. For
instance they play an important role in some  versions of string or branes
theories, and constitute presently the only cosmological example of
the holographic conjecture.

Recent calculations \cite{MIZO}  suggested that the \dS ~solution play an universal
role as an ``osculating" manifold for space-time.

Since the beginning of
the eighties,  the de Sitter space has been
considered as a key model in inflationary cosmological scenario
\cite{[LI]}, where it is assumed that the cosmic dynamics was
dominated by a term acting like a cosmological constant.
More recently, observations on far high redshift
supernovae \cite{[PER]}, on galaxy clusters \cite{[PH]}, and on
cosmic microwave background radiation \cite{[SAP]} suggested an
accelerating universe. Again, this can be explained in a satisfactory way  with such a
term.
This current  ``inflation" is based on (increasingly  reliable) current observations. The other one is of totally dynamical nature and is still subject to controversies. This can be summarized in decomposing $\Lambda$ into $\Lambda_{\mathrm{bare}}$ and $\Lambda_{\mathrm{vacuum}}$, \emph{i.e.}, into a bare cosmological background and an extra term which is of quantum  origin, the latter assuming large enough values for  justifying  inflation scenario. Also, it is obvious that $\Lambda$ is not thought as the unique responsible for the complete history of the growth of the universe. Other matter entities ($\rho_{\mathrm{matter}}$, $\rho_{\mathrm{rad}}$, ...) are important in different epochs. 
In our paper,  we  intend to explain  at length what could be the consequences of having a constant, non-null,  $\Lambda$, whatever its origin,  on our way of considering masses. 

On the fundamental level,   matter and energy are of quantum nature.
But the usual quantum field theory is designed in Minkowski space-time. Most of  the theoretical and
observational arguments privileging   a   de Sitter-like universe
plead in favour of setting up a rigorous quantum field theory in \dS  space-time, or at least exploring specific features  
which could show up  in such a framework and which would not have any counterpart in the flat curvature limit.
Fortunately, the symmetry properties of dS universes may allow the
construction of such a theory. For recent works on this subject, see for instance \cite{[BEM],[GT2]} and references therein.
We should also note that the study of de Sitter   space-time offers a specific interest because of   the regularization
opportunity afforded by the curvature parameter as a ``natural''
cutoff for infrared or other divergences.
On the other hand, we should be also aware that  some of our most familiar concepts like time (see for instance the ambiguity in choosing the static coordinate time versus the conformal time), rest mass, energy,
momentum, etc, disappear, or at least need radical modifications in
de Sitterian relativity, as we will comment later on.

\subsection{Massive or Massless?}

In Minkowski, the concept of (rest) mass originates in the ubiquitous law of conservation of energy, a direct consequence of the Poincar\'e symmetry. As soon as we deal with de Sitter or anti de Sitter background, this concept of mass should be totally reconsidered.  In particular, one may expect to lose a precise distinction between ``massive'' and ``massless''. So, we should look for other properties, e.g. existence or violation of conformal invariance,  of some gauge invariance, in view of extending concepts about mass inherited from minkowskian physics.

The main purpose of the present paper is to bring attention, in the framework of EGR-1-2 and group representation theory, on the way we usually consider the vanishing of graviton mass. We compare the content of the linearized perturbation  of the fundamental states of EGR-1 and EGR-2,  respectively Minkowski and de Sitter/anti de Sitter space-times, with the content of spin-2 ``massive'' and ``massless'' standard wave equations in these three possible backgrounds. The issue of this analysis is that, although the concept of mass is non ambiguous in the case of Minkowski, it is no longer true for the other cases. In consequence, we shall see that giving a mass or not to graviton might  depend on the fundamental nature of space-time. Moreover, we will not restrict our analysis to the spin-2 case only. We will examine the arbitrary spin case within a group theoretical framework. This point of view leads us to adopt a definition of mass in dS/AdS which has been recently proposed by Garidi in \cite{GAR1}. The Garidi definition advantageously gives sense to terms like ``massive''  and ``massless''  fields in dS/AdS with regard to their minkowskian counterparts obtained through group contraction procedures.

The organization of this paper is as follows. In Sections \ref{cosmograv1}, \ref{cosmogravC} and \ref{cosmogravD} we examine the question of the graviton mass with the hypothesis of a non-zero cosmological constant $\Lambda$ by revisiting  some of the standard arguments against the idea of a massive graviton  from a minkowskian point of view, namely arguments based on the number of degrees of freedom (Section \ref{cosmograv1}) and on the long-range of the gravitational interaction (Sections \ref{cosmogravC} and \ref{cosmogravD}). In Section V we carry out a short numerology analysis on two types of scales involving $\Lambda$. One is from quantum origin and  rests upon the existence, for any non zero mass $m$, of the dimensionless parameter $\vth \equiv \vth_m =:  \frac{\hbar \sqrt{\vert \Lambda} \vert}{\sqrt{3}mc}$ which characterizes any minkowskian asymptotic dS/AdS physics for massive quantum systems. The  other one involves the Newton gravitational constant. The issue concerning a massive graviton is expressed in terms of a ratio similar to that one existing in the so-called \emph{cosmological constant puzzle}. In Sections VI and VII is reconsidered the spin 2 equation in a curved background. In the particular dS or AdS backgrounds, it is found that a fine tuning minkowskian mass is necessary in order to reduce to 2 the number of degrees of freedom. The next sections are devoted to the group theoretical approach. Section VIII is a review of the group material: geometries, symmetries and classification of representations of the dS and AdS groups. Since we are mainly concerned by the relations between minkowskian masses and dS and AdS ``masses'', we describe in a comprehensive way in Sections IX and X the mathematical process of contractions \emph{dS \& AdS $\to$ Poincar\'e} and \emph{dS \& AdS $\to$ their nonrelativistic counterparts ``Newton$_+$'' and ``Newton$_-$''}. We will insist on the fact that there exist fields in dS/AdS, like the ubiquitous (e.g. in inflation scenario)  ``minimally coupled massless'' field, which do not have any minkowskian counterpart.
 Finally, we reexamine within this contraction framework the question of mass in dS and AdS space-times and examine with more details the nature of the parameter $\vth_m$. In this respect, we  underline the important fact that, due to the infinitesimal current value of $\Lambda$, $\vth_m$ is totally negligible for all known massive elementary particles: no dS effect is perceptible in LHC experiments! On the other hand, any theory giving an infinitesimal, yet non-zero, mass, to  photon or graviton or to other massless gauge field, or yet based on large values of $\Lambda$ should take in consideration some physical $\Lambda$ effect as soon as $\vth_m$ gets closer to 1.

\setcounter{paragraph}{0}
\section{Standard arguments against associating the graviton
mass to $\Lambda$}
 \label{cosmograv1}

The standard  reasons invoked in order  not to
associate the cosmological constant $\Lambda$ to the mass of the
graviton are based on two arguments:
\begin{itemize}
\item{The massive graviton has extra degrees of freedom;}
\item{The Yukawa analysis shows that the range of gravity should
be cut off at distances greater than its Compton wave-length
$l_g \equiv  m_{g}^{-1}$}  (in units $c=1 =\hbar$).
\end{itemize}
These arguments are correct in the
specific minskowkian framework only. They can be completely bypassed  if the
fundamental state of the background  geometry is not Minkowski,
but instead is of a de-Sitter/anti-de-Sitter type. In other words,
we will show that both arguments are not robust when passing to
equations of EGR-2. To be more explicit, let us state that
\begin{itemize}
\item{the existence of a graviton mass term proportional to $\sqrt{\vert\Lambda \vert}$ in a de Sitter
background does not produce extra degrees of freedom;} 
\item{the
standard Yukawa analysis does not apply. In other words, the
corresponding range of gravity is not cut off at distances greater
than the Compton wave-length $m_{g}^{-1}$ of the graviton.}
\end{itemize}
Let us analyse each one of these arguments separately.

\subsection{Spontaneous broken symmetry and the graviton mass}
\label{cosmogravA}

Goldstone boson mechanism has been used in gauge theories as a suitable procedure
 to generate mass to the gauge fields. This was used with
success in the standard $SU(2)\times U(1)$ electroweak
unification to provide mass for the vectorial bosons that mediate
weak processes. The idea goes back to the possibility of having a
scalar field $\Phi$ the dynamics of which is described by a
Lagrangian which, besides the kinetic term, has a self-interaction
described by a potential $V(\Phi).$ If there exists a
solution $\Phi = constant$ that extremizes the corresponding
potential, this can provide a constant term in the interaction with
a massless gauge field (like electromagnetic vector field). This term is precisely  interpreted as the mass of the
gauge field. Consequently, the associated
energy-momentum tensor takes a form proportional to the metric
tensor, that is, $ T_{\mu\nu} = V_{0} g_{\mu\nu}.$ On the other hand, in the
case of gravity, a similar mechanism is not interpreted as a mass
term. Instead of giving mass to the graviton, this term is
interpreted as a change of the background fundamental geometry.

One more comment is of importance at this point. It has
been pointed out that any classical nonlinear field theory can
act in the same way and provide a ``cosmological term'', no matter  it be a scalar or not. Indeed, take for
instance a nonlinear electrodynamic theory described by the
Lagrangian $L =L(F)$ where $F \equiv F_{\mu\nu} F^{\mu\nu}$ the
dynamics is given by
\begin{equation}
\left( L_{F} \, F^{\mu\nu} \right) _{;\nu} =0, \label{27nov10}
\end{equation}
where $L_F =: \mathrm{d}L/\mathrm{d}F.$
The corresponding energy-momentum tensor is given by
\begin{equation}
T_{\mu\nu} = L_{F} \, F_{\mu\alpha} \, F^{\alpha}{}_{\nu} - L \,
g_{\mu\nu} \label{27nov11}
\end{equation}
Thus, the state in which $L_{F} $ vanishes for the constant
solution $F = F_{0}$ yields a term proportional to the metric
tensor which mimics the cosmological constant.

\setcounter{paragraph}{0}
\subsection{The graviton mass and the cosmological constant ratio puzzle.
Observational limits}
\label{cosmogravB}
Before going into the details of the calculations
let us examine for a while  what we can learn from
the cosmological equations. In the standard FRW framework, the
metric assumes the form
\begin{equation}
ds^{2} = dt^{2} - a^{2}(t) \, d\sigma^{2}. \label{noite1}
\end{equation}
The basic equation that provides the main constraint on the
evolution of the universe reads as:
\begin{equation}
\frac{1}{3} \, \Theta^{2} = \frac{\Gamma}{a^{4}} + \frac{M}{a^{3}}
+ \frac{\Phi_{0}}{a^{n}} - 3 \, \frac{\epsilon}{a^{2}} + \Lambda.
\label{27nov13}
\end{equation}
in which $\Theta \equiv 3 \, \dot{a}/a$ is the expansion factor
(the Hubble ``constant"), $ \Gamma, M$ and $\Phi_{0}$ are
constants; $n
> 0$; $\epsilon$ is the value of the 3-section and $a(t)$ is the
scale factor. We conclude from this equation that at the limit $
a \rightarrow \infty$ the origin of the spacetime curvature is mainly due to
 the $\Lambda$-term:
 its importance is such that it will dominate over all other
forms of energy in the far future.

 The main question one faces to in the
identification of $\Lambda$ with the graviton mass can be
formulated as follows:
\begin{itemize}
\item{How to conciliate a possible graviton mass with the observational fact that
gravity is a long range force?}
\end{itemize}
This question has been considered many times in the literature
(see for instance \cite{goldhaber} and related papers quoted
therein). These authors put the question into the following
perspective. Within the realm of Theory EGR-1, using an analysis based on analogy 
with Yukawa's interpretation of massive exchanging particles, it is possible to establish from observation
the upper bound:
$$ m_{g}^{2} < 2.10^{-29} \, h_{0}^{-1} \mathrm{eV},$$
where $h_{0}$ is the normalized Hubble constant.
This yields for the associated Compton length the lower bound
$$ l_{g} > 10^{24}  \mathrm{cm}.$$
This should be compared  with the cosmological limit:
$$ \vert \Lambda \vert^{-\frac{1}{2}} \approx 10^{28} \mathrm{cm}.$$
There follows the relation
\begin{equation}
l_{g} \, \vert \Lambda\vert ^{\frac{1}{2}} \approx 10^{-4}. \label{27nov12}
\end{equation}
What is the reason for this relation? Why the current limits set by
cosmological observations yield  so near values for two
apparently uncorrelated quantities? In the present paper we
present arguments that provide an explanation for such a quite coincidence. Moreover, we
suggest a prediction that future observations will improve the coincidence
 in such a way that one will eventually arrive at the
relation $l_{g} \, \vert \Lambda\vert ^{\frac{1}{2}} \approx 1.$

\setcounter{paragraph}{0}
\section{The Yukawa potential}
\label{cosmogravC}
In order to set the limits pointed out in the previous section, as is done for instance 
in the latest issue of the Particle Data Properties \cite{pada}, a scenario is
described in which the basic framework is provided by the so-called Yukawa potential. This idea, conceived and meaningful in
flat Minkowski background, states that the massive gravitational
field has an  effect by diminishing the effective gravitational field through the modified potential
\begin{equation}
 V(r) = \frac{\exp{(-\mu r)}}{r} \label{c1}
\end{equation}
which satisfies the massive  Laplace equation
\begin{equation}
\nabla^{2} V(r) + \mu^{2} \, V(r) = 0. \label{y1}
\end{equation}
From (\ref{c1}) it follows the cut-off of the interaction on
distances that depends uniquely on the mass value. Should one
applies this idea irrespectively of the structure of the
background geometry? Or, in other words, could this same reasoning
be applied both to the dynamics controlled by EGR-1 or EGR-2
equations of motion? The answer, we anticipate it, is no. Let us
prove this in the simplest case of a massive scalar field in a (anti-)
de Sitter background. But before this, let us
ponder why one should be prepared to accept this modification. The
reason comes from the simple fact that contrary to the minkowskian
case, in de Sitter geometry the constant curvature $\Lambda$ is a 
quantity which is intrinsic to the geometry and  which  acts as a fundamental length. Thus one
should expect that the main properties of the field do not depend on
the absolute value of the mass of the test-field but on only the ratio
of the mass of the field to a  fundamental mass constructed from $\Lambda$ or,
equivalently, on the ratio of their associated Compton
wavelengths.

\setcounter{paragraph}{0}
\section{The case against the gravitational Yukawa potential}
\label{cosmogravD}
In order to simplify our description we will treat the case of a
scalar field. Let us set the background geometry and treat both
cases according whether $\Lambda$ be positive or negative. The metric reads in  \emph{static} coordinates as:
\begin{equation}
  ds^{2} =(1 - \frac{\Lambda}{3} \, r^{2}) dt^{2} - (1 -
\frac{\Lambda}{3} \, r^{2})^{-1} \,dr^{2} - r^{2} (d \theta^{2} +
\sin^{2}{\theta} d \varphi^{2}) . \label{y1}
\end{equation}
Note that these coordinates are not global in the dS case, since there exists there a horizon.
In the minimal coupling case (``massless minimally coupled field'') the equation of motion reads 
\begin{equation}
\Box \varphi = 0, \label{27nov31}
\end{equation}
where $\Box $ is the Laplace-Beltrami operator in coordinates (\ref{y1}).
Let us first consider the AdS case in which $\Lambda < 0$.
 For the static and spherically symmetric case (\ref{27nov31}) reduces to
\begin{equation}
\left( ( 1 - \frac{\Lambda}{3} \, r^{2}) \, y^{'} \right)^{'} +
\frac{2}{3} \, \Lambda \, y = 0, \label{27nov32}
\end{equation}
in which we have set $\varphi = y(r)/r.$ Besides the trivial solution $\varphi =$ constant, the  regular solution which  vanishes at $r \to \infty$ is
immediately found to be proportional to
\begin{equation}
\varphi (r) = \frac{1}{r} + B \arctan{(Br)} - B \frac{\pi}{2},
\label{27nov33}
\end{equation}
where $ B^{2} \equiv - \, \frac{\Lambda}{3}.$ For the case of a
positive cosmological constant, the domain of validity of the coordinate system is restrained to $ 0  \leq  r^{2} < 3/\Lambda
=: C^{-2}.$ In this case, no  nontrivial solution regular at the boundary $Cr = 1$ exists, since the counterpart of (\ref{27nov33}) reads

\begin{equation}
\varphi (r) =  \frac{1}{r} - C\tanh^{-1}(Cr).
\label{4JAN1}
\end{equation}

Nevertheless, another choice of coordinate sytem (\emph{e.g.} global conformal coordinates) allows to circumvent this (apparent?) problem.

 The question now is: to what extent will the  result  obtained in the AdS case be modified if
the scalar field is massive? The equation in this case reduces to
a Legendre type
\begin{equation}
\left( 1 - \frac{\Lambda}{3}\, r^{2} \right) y^{''} - \frac{2}{3}
\, \Lambda r y^{'} - (M^{2} - \frac{2}{3} \Lambda) y = 0.
\label{27nov34}
\end{equation}
A numerical analysis of the solutions shows that in the case in
which the mass of the scalar field is lower than twice
$\vert\Lambda\vert^{-\frac{1}{2}}$ the field behavior is very similar to the
massless case, and it decays more strongly for the case in which
$M > 2 \vert\Lambda\vert^{-\frac{1}{2}}.$ Hence one should
suspect that the decay of the field with the distance depends on
the ratio between the Compton wavelength of the field and
$\vert\Lambda\vert^{\frac{1}{2}}.$

\setcounter{paragraph}{0}
\section{Numerology}

From a QFT point of view  the cosmological
constant is usually  identified to the vacuum energy density
$\rho_{Pl}.$ Using this assumption, the ``natural  value of this
constant" is
\begin{equation}
\rho_{Pl} \approx M_{Pl}^{4}\, \frac{c^{3}}{\hbar^{3}}
\label{24nov}
\end{equation}
where $M_{Pl}$ is the Planck mass.

Although there is a common belief that this quantity should play
an important role in the quantum gravitational world, one should
not take it as independent of a certain bias concerning the basic
numeric ingredients at this level. None the less, let us follow a
similar procedure in order to examine the consequences one could
be led to if one accepts the idea that  the cosmological constant $\Lambda$ is a
true independent fundamental number in our universe from which the
value of the graviton mass is related.

\subsection*{Two masses?} 
Let us ask the following question:
what is the value of the mass that one
should expect to be associated to $\Lambda ?$ In the quantum
context, the natural quantity should be constructed with three
basic ingredients: $\Lambda$, Planck constant $\hbar$ and light
speed $c.$ Given a non-zero $\Lambda$ and any non zero mass $m$, one can deal with the  dimensionless parameter $\vth \equiv \vth_m =:  \frac{\hbar \sqrt{\vert \Lambda} \vert}{\sqrt{3}mc}$. We will give more details on this parameter in Section XI, while considering  flat limit of dS/AdS framework. The factor $\frac{1}{\sqrt{3}}$ is irrelevant but is put here for convenience.

When $\vth$ is of the order of the unity, we  are  led, up to a numerical factor, irrelevant also for the present  discussion,  to the formula
\begin{equation}
m_{\Lambda} = \frac{\hbar \, \sqrt{\vert \Lambda \vert }}{c}. \label{4MAr3}
\end{equation}
Note also that this is the unique mass provided by any field theory using a
quantity which has the dimension of length. 

Nevertheless, as it has
been emphasized many times, the graviton is not just one of those
particles that happens to exist in any metrical structure: it is
special. This is because the graviton is intrinsically related to
the metrical structure of space-time. In this vein, there exists
another quantity that has dimension of mass constructed with $
\Lambda.$ This second mass should contain three basic ingredients
too, but should not be dependent on the quantum world: it should  instead
 exhibit its gravity dependence. This means that the second
quantity, call it $M_{\Lambda}$, is constructed with $\Lambda$, the Newton
constant $G_{N}$ and the light velocity $c.$ This yields, up to a
numerical factor, the expression
\begin{equation}
M_{\Lambda} = \frac{c^{2}}{G_{N} \, \sqrt{\vert \Lambda \vert }}. \label{4MAr4}
\end{equation}
Before continuing let us  ask about the
meaning that one should attribute to this $M_{g}.$ In order to
clarify this let us rewrite it in an equivalent form as
\begin{equation}
M_{\Lambda} = \frac{\Lambda \, c^{4}}{G_{N}} \,
\frac{1}{\sqrt{\vert \Lambda \vert^{3}}} \, \frac{1}{c^{2}}. \label{4MAr5}
\end{equation}
This contains three separate factors. The first one represents the (gravitational)
energy density generated by $\Lambda$, say $\rho_{\Lambda}c^2;$ the second term
($\vert \Lambda\vert^{-\frac{3}{2}}$) is the total volume of the universe
restricted to its horizon $H_{0}^{2} \approx \vert\Lambda\vert^{-1}$ and the
last term is just there to convert the total energy into a mass. Thus
 we are almost constrained to interpret  $M_{\Lambda}$ as the total mass of all existing
gravitons in the observable universe. It then follows that if we
 write
\begin{equation}
M_{\Lambda} = N_{g} \, m_{\Lambda}, \label{4Mar6}
\end{equation}
then $N_{g}$ is to be interpreted as the total number of gravitons
contained in the observable horizon.

In this respect, there have been some arguments concerning the existence of a residue of gravitational waves (see for instance \cite{grisch} for some consequences of relic gravitational waves in cosmic microwave background radiation and for aÊlistÊ of references) which could be associated to a huge number of gravitons in equilibrium in a similar way as for a background ofÊ neutrinos and photons.

An unexpected result appears when we evaluate this quantity in our
actual universe: it is exactly the same number that appears in the
standard cosmological constant puzzle. Indeed, from the above
expressions we obtain:
\begin{equation}
 \frac{ \rho_{Pl}}{\rho_{\Lambda}} \approx \frac{c^{7}}{\hbar \, G_{N}^{2}}
\, \frac{G_{N}}{c^{4} \, \vert \Lambda \vert} ,\label{4Mar7}
\end{equation}
and we get the same quantity for the ratio of the two masses
\begin{equation}
 N_{g} = \frac{c^{3}}{\hbar \, G_{N}
\, \vert \Lambda \vert} .\label{4Mar7bis}
\end{equation}
Consequently,
\begin{equation}
\frac{\rho_{Pl}}{\rho_{\Lambda}} \approx N_{g}\approx 10^{120}.
 \label{4Mar8}
\end{equation}
This analysis implies that the traditional cosmological constant
problem and $N_{g}$ have the same common origin and led us to
suggest that the value of $\rho_{Pl}/\rho_{\Lambda}$ is so large because
there is a huge quantity of massive gravitons, with $m_{\Lambda} \approx
\sqrt{\vert \Lambda\vert },$ in the observable universe.

\setcounter{paragraph}{0}
\section{Equation of spin-2 in curved background}
The passage of the spin-2 field equation   from Minkowski
spacetime to arbitrary curved riemannian manifold presents
ambiguities due to the presence of second order derivatives of the
rank two symmetric tensor $\varphi_{\mu\nu}$ that is used in
the so called Einstein-frame (see for instance
\cite{Aragone-Deser-2}). These ambiguities disappear when we pass to
the Fierz frame representation that deals with the three index
tensor 
$$F_{\alpha\mu\nu} = \frac{1}{2}\left\lbrack \varphi_{\nu \alpha,\mu}- \varphi_{\nu \mu,\alpha} + F_{\alpha} \eta_{\mu\nu} - F_{\mu} \eta_{\alpha\nu}\right\rbrack,$$ 
where $F_{\alpha} = F_{\alpha\mu\nu}\eta^{\mu\nu} = \varphi_{,\alpha} - {\varphi_{\alpha}{}^{\mu}}_{,\mu}.$
This was shown in \cite{Novello-Neves}.
There results  a unique form of minimal coupling, free
of ambiguities. Let us define from $\varphi_{\mu\nu}$ two auxiliary fields $G^{(I)}{}_{\mu \nu}$
and $G^{(II)}{}_{\mu \nu }$ through the expressions:
\begin{align}
\nonumber 2 \, G^{(I)}{}_{\mu \nu }&\equiv Ê\\Ê
\Box \,\varphi _{\mu \nu
}-\varphi_{\epsilon(\mu ;\nu )}{}^{;\epsilon }&+\varphi _{;\mu \nu
}-\eta _{\mu \nu }\,\left( \Box \varphi -\varphi ^{\alpha \beta
}{}_{;\alpha \beta }\right) , \label{28nov1}\\
\nonumber 2 \, G^{(II)}{}_{\mu \nu }&\equiv  \\
\Box \,\varphi _{\mu \nu
}-\varphi_{\epsilon(\mu}{}^{;\epsilon}{} _{;\nu)}&+\varphi _{;\mu
\nu }-\eta _{\mu \nu }\,\left( \Box \varphi -\varphi ^{\alpha
\beta }{}_{;\alpha \beta }\right). \label{28nov2}
\end{align}
These objects differ only in the order of the second derivative in the second term on the r.h.s. of the above equations.
The equation of motion free of ambiguities concerns the tensor field
\begin{equation}
\widehat{G}_{\mu \nu } \equiv \frac{1}{2} \, \left( G^{(I)}{}_{\mu
\nu } + G^{(II)}{}_{\mu \nu } \right) \label{28nov3}
\end{equation}
and is given by
\begin{equation}
\widehat{G}_{\mu \nu } + \frac{1}{2} \,m^{2} \, (\varphi_{\mu \nu}
-\varphi g_{\mu\nu})= 0. \label{28nov4}
\end{equation}

\subsection*{The fine-tuning mass}
 The field $\varphi_{\mu \nu}$ has five degrees of
freedom. In the case of  flat minkowskian background the massless field has two
degrees of freedom. This is a consequence of the invariance of the
action under the gauge transformation
\begin{equation}
\varphi_{\mu\nu} \rightarrow \varphi_{\mu\nu} + \delta
\varphi_{\mu\nu} \label{30nov1}
\end{equation}
in which
\begin{equation}
\delta \varphi_{\mu\nu} = \xi_{\mu , \nu} + \xi_{\nu , \mu}
\label{30nov2}
\end{equation}

In the case of de Sitter (dS or AdS) a similar analysis can be made.
However, in a quite unexpected way, something puzzling occurs
which is at the basis of the discussion concerning the value of
the mass for the graviton. Let us summarize the
situation. The action, in the Fierz-frame, is given by
\begin{align}
\nonumber S &= \\
& \frac{1}{4} \, \int \sqrt{-g} \left[ A - B - \frac{m^{2}}{2}
\, \left(\varphi_{\mu\nu} \, \varphi^{\mu\nu} - \varphi^{2}
\right) \right] d^{4}x, \label{30nov3}
\end{align}
where $A \equiv F_{\alpha\beta\mu} \, F^{\alpha\beta\mu}$ and $B
\equiv F_{\mu} \, F^{\mu}.$ With the curved counterpart $\delta \varphi_{\mu \nu} = \xi_{\mu ; \nu} + \xi_{\nu ; \mu}$ of the  transformation
(\ref{30nov2})   the action (\ref{30nov3}) changes to
\begin{equation}
\delta S = \frac{1}{2} \, \int \sqrt{-g} \left( Z^{\mu} - m^{2}
F^{\mu} \right) \xi_{\mu} d^{4}x, \label{30nov5}
\end{equation}
in which we have defined
$Z^{\mu}$ as the divergence
$$ Z^{\mu} \equiv 2 \widehat{G}^{\mu\nu}{}_{; \nu}.$$

We note that for general curved riemannian background this action
will not be invariant under such a transformation. However, when
the geometry is that of a de Sitter/anti de Sitter spacetime the quantity $Z^{\mu}$ has
the value $Z^{\mu} = - \frac{2}{3} \Lambda F^{\nu}.$
Hence, for  the exceptional case in which the mass of the
spin-2 field has the special value 
\begin{equation}
\label{gravmass}
m^{2} = - \frac{2}{3}
\Lambda,
\end{equation}
 the action displays a gauge invariance and the field has
only two degrees of freedom. The surprising feature of this result
is that the action invariance, in a de Sitter/anti de Sitter
background,  does not occur for the massless field, but instead, for
the field that satisfies \ref{gravmass}. In this
case the field has only two degrees of freedom, and for any other
value of the mass, it has five. Now it raises the question: does that very special value of the minskowkian mass represent any known particle (in the AdS case) or tachyon (in the dS case)? We shall see in the next section that  a
particle  presenting such a special value of the mass (in the minkowskian sense) 
 should be identified as the graviton, that is, the
tensorial wave perturbation of de Sitter/anti de Sitter background
geometry.

\setcounter{paragraph}{0}
\section{Einstein spaces}

By definition, an Einstein space is such that its geometry
satisfies the equation EGR-2 without matter:
\begin{equation}
R_{\mu\nu} - \Lambda g_{\mu\nu} = 0. \label{27nov1}
\end{equation}
It includes, in particular the de Sitter and anti de Sitter
geometries. The riemannian curvature $R_{\alpha\beta\mu\nu}$ is
written in terms of the Weyl conformal tensor and
$g_{\alpha\beta\mu\nu} \equiv g_{\alpha\mu} g_{\beta\nu}
-g_{\alpha\nu} g_{\beta\mu},$ that is:
\begin{equation}
R_{\alpha\beta\mu\nu} =  W_{\alpha\beta\mu\nu} + \frac{\Lambda}{3}
\, g_{\alpha\beta\mu\nu}. \label{27nov2}
\end{equation}
The question we face to here is the following: what is the relation of the
equation of motion of a gravitational perturbation in this theory
and how does it compare  with the  spin-2  equation (\ref{28nov4}) presented in the
previous section?

\bigskip
Let us write the equation (\ref{27nov1}) under the form
\begin{equation}
G_{\mu \nu }+\Lambda \text{ }g_{\mu \nu }=0, \label{4.1}
\end{equation}
Let us perturb this equation around the de Sitter solution
${g}_{\mu\nu,}$ where the Weyl tensor vanishes. Using the notation
$\delta g_{\mu \nu }= h_{\mu \nu },$ the equation for the
perturbed field is provided by
\begin{equation}
\delta G_{\mu \nu }+\Lambda \, h_{\mu \nu }=0, \label{4.3}
\end{equation}
where $\delta G_{\mu \nu }$ is the perturbation of the Einstein
tensor. A direct calculation yields
\begin{equation}
\delta G_{\mu \nu }=G^{(I)}{}_{\mu \nu} +A_{\mu \nu },
\end{equation}
\label{4.4} where
\begin{equation}
A_{\mu \nu }=\frac{1}{2}\left( g_{\mu \nu } \, R^{\alpha \beta} \,
h_{\alpha \beta}- R \, h_{\mu \nu }\right) . \label{4.5}
\end{equation}
Manipulating equations (\ref{28nov1}), (\ref{28nov2}) and
(\ref{28nov3})
 we arrive at
\begin{equation}
G^{(I)}{}_{\mu \nu }=\widehat{G}_{\mu
\nu}-\frac{1}{2}R_{\mu\alpha\nu \beta }\,h^{\alpha
\beta}+\frac{1}{4}\,R_{\alpha (\mu}\, h_{\nu )}{}^{\alpha },
\label{4.6}
\end{equation}
and using the property that
$$ R_{\alpha\beta\mu\nu} = \frac{\Lambda}{3} \left( g_{\alpha\mu}
g_{\beta\nu} - g_{\alpha\nu} g_{\beta\mu} \right), $$ Equation
(\ref{4.3}) becomes
\begin{equation}
\widehat{G}_{\mu \nu }-\frac{\Lambda }{3}\left( h_{\mu \nu}-h \,
g_{\mu \nu }\right) =0, \label{4.10}
\end{equation}
where $h = h_{\mu\nu} \, g^{\mu\nu}.$

Comparing with equation (\ref{28nov4}), one recognizes that the
graviton has  the ``mass'' 
\begin{equation}
\label{gravmass1}
m_{g}^{2} = - \frac{2}{3}\Lambda,
\end{equation} in agreement with (\ref{gravmass}), and this mass is precisely of the same order as the mass $m_{\Lambda}$ we have introduced in (\ref{4MAr3}).

With a negative curvature (AdS), $m_g$ can be interpreted as a graviton minkowskian mass. However, in the case of positive curvature (dS), such an identification raises the dilemma to face or not to the question of existence of a ``tachyonic graviton'' and so could create serious interpretative difficulties. Actually, this dilemma  is only apparent
because  another interpretation of Equation (\ref{gravmass1}) is possible and consistent within the framework of the de Sitter group representation (EGR-2 point of view), which we carry out in the next sections.  Indeed, Eqs. (\ref{28nov4}) and (\ref{4.10}) are wave equations for a spin 2 dS elementary system which propagates on the light cone and has two degrees of freedom only, exactly like a massless minkowskian graviton would have if there were no curvature. A massless graviton in de Sitter seems to appear as a tachyonic particle in Minkowski!

\section{(Anti-) de Sitter geometries and (quantum) kinematics}

We now start with the group theoretical material needed for explaining the EGR-2 point of view and attempting to give a firm base to our critical examination of the problem of the graviton mass.
Of course, we will have to modify our way of reasoning  about notions like \emph{mass}, \emph{massive},  \emph{massless} or even \emph{energy}. As was announced in the introduction,  we will not restrict our analysis to the spin-2 case only. We will rather examine the arbitrary spin case in order to offer a comprehensive overview of the basic dS/AdS mathematical material.
\subsection{Hyperboloids}
\setcounter{paragraph}{0}

As was already stated in the introduction,  the de Sitter (resp. anti de Sitter) metric is the unique solution of the cosmological  vacuum
Einstein's equation with positive (resp. negative) cosmological constant $\Lambda=3~\vka^{2}$.
\bqn
\label{eins}
R_{\mu \nu} - \frac{1}{2} R g_{\mu \nu} + \Lambda g_{\mu \nu} = 0,\\
\nonumber R= R_{\mu \nu}g^{\mu \nu}= 4 \Lambda \equiv 12\vka^2. \eqn

\paragraph{de Sitter geometry}

The corresponding de Sitter space is conveniently described  as an
one-sheeted hyperboloid embedded in a five-dimensional Minkowski
space (the bulk):
\begin{align}
\nonumber
    M_{dS} \equiv& \{x \in \R^5 ;~x^2=\eta_{\alpha\beta}~ x^\alpha
x^\beta =-\vka^{-2}\},   \\
\label{dshyp}    & \alpha,\beta=0,1,2,3,4,
\end{align}
where $\eta_{\alpha\beta}=$diag$(1,-1,-1,-1,-1)$.
We can introduce for instance the global coordinates $\tau \in \R, \vec{n} \in S^2, \alpha \in [0,\pi]$:
 \begin{align}
 \label{coordh}
\nonumber x:=& (x^{0} ,
\vec{x}=(x^{1},x^{2},x^{3}), x^4)\\
\nonumber =& (\vka^{-1}\sinh(\vka c t), \, \vka^{-1}\cosh(\vka c t)\sin(\vka r)\vec{n}, \\
& \vka^{-1}\cosh(\vka ct )\cos(\vka r)).
\end{align}

\paragraph{anti de Sitter geometry}

The anti de Sitter space can be viewed as an
one-sheeted hyperboloid embedded in another  five-dimensional
space with different metric:

\begin{align}
\nonumber
    M_{AdS} \equiv& \{x \in \R^5 ;~x^2=\eta_{\alpha\beta}~ x^\alpha
x^\beta =\vka^{-2}\},   \\
\label{adshyp}    & \alpha,\beta=0,1,2,3,5,
\end{align}
where $\eta_{\alpha\beta}=$diag$(1,-1,-1,-1,1)$.
Global coordinates $\tau \in [0, 2 \pi), r \in [0, \infty), \vec{n} \in S^2$ are defined  by:
 \begin{align}
 \label{adscoordh}
\nonumber x:=& (x^{0},
\vec{x}=(x^{1},\, x^{2}, x^{3}),\,  x^5)\\
\nonumber =& (\vka^{-1}\cosh{(\vka r)}\,\sin(\vka c t), \vka^{-1}\sinh(\vka r)\, \vec{n},\\
& x^{5}=\vka^{-1}\cosh(\vka r )~\cos(\vka c t).
\end{align}

\setcounter{paragraph}{0}
\subsection{The de Sitter group, its unitary irreducible representations, and their physical interpretation }

The de Sitter relativity group is $G=SO_{0}(1,4)$, i.e. the component
connected to the identity of the ten-dimensional pseudo-orthogonal
group $SO(1,4)$. A familiar realization of the Lie algebra is that
one generated by the ten Killing vectors
\beq
\label{kil}
K_{\alpha \beta} = x_{\alpha}\partial_{\beta} -
x_{\beta}\partial_{\alpha}. \eeq
It is worthy to notice that there is no globally time-like Killing
vector in de Sitter, the adjective time-like (resp. space-like)
referring to the Lorentzian   four-dimensional
 metric induced by that of the bulk.

The universal covering of the de Sitter group is the symplectic
$Sp(2,2)$ group, which  is needed when dealing with half-integer spins.

Specific quantization procedures  applied to  classical phase
spaces viewed as co-adjoint orbits of the group lead to their quantum counterparts, namely the quantum
elementary systems associated in a biunivocal way to  the UIR's of
the de Sitter group $Sp(2,2)$. Let us give a complete classification
of the latter, following the work by Dixmier \cite{[DIX]}. We recall that the ten Killing vectors (\ref{kil})
can be represented as (essentially) self-adjoint operators in Hilbert
space of (spinor-)tensor valued functions on $M_{dS}$, square integrable
with respect to some invariant inner product, more precisely of the
Klein-Gordon type. These operators take the form
  \beq
  \label{genrep}
K_{\al \be} \lga L_{\al \be} = M_{\al \be} + S_{\al \be},
  \eeq
where the orbital part is $M_{\al \be}=-i(x_{\al}\partial_{\be} -
x_{\be}\partial_{\al})$ and the spinorial part $S_{\al \be}$ acts on
the indices of functions in a certain permutational way. Like for the UIR of the Poincar\'e group, there are
two Casimir operators, the eigenvalues of which determine completely
the UIR's. They respectively read:
\beq
\label{cas1}
Q^{(1)} = - \frac{1}{2} L_{\al \be}L^{\al \be}, \eeq
with eigenvalues
\begin{equation}
\label{eigends1}
\langle Q^{(1)} \rangle_{\mathrm{dS}} =  -p(p+1) - (q+1)(q-2),
\end{equation} and
\begin{equation}
\label{cas2}
Q^{(2)} = - W_{\al}
W^{\al}, \ W_{\al} = - \frac{1}{8}\epsilon_{\al \be \ga \delta \eta}
L^{\be \ga}L^{\delta \eta},    \end{equation}
with eigenvalues
\begin{equation}
\label{eigends2}
\langle Q^{(2)} \rangle_{\mathrm{dS}} =  -p(p+1)q(q-1).
\end{equation}

Therefore, one must distinguish between
\bei
\item{\bf The discrete series} $\Pi^{\pm}_{p,q}$, \\
defined by $p$ and $q$ having integer or half-integer values, $p \ge q$,
$q$ having a spin meaning.
Here, we must again distinguish between
\bei
\item {\it The scalar case} $\Pi_{p,0}$, $p=1,2, \cdots$;
\item {\it
The spinorial case} $\Pi^{\pm}_{p,q}$, $q>0$, $p= \frac{1}{2}, 1,
\frac{3}{2}, 2, \cdots$, $q=p, p-1, \cdots, 1$ or $\frac{1}{2}$ \ei

\item {\bf The principal and complementary series} $\NU_{p,\sigma}$,
\\ where $p$ has a spin meaning.
We put $\sigma =q~(1-q)$ which gives $q = \frac{1}{2}\left( 1 \pm \sqrt{1 - 4 \sigma^2} \right)$.

Like in the above, one distinguishes
between
\bei
\item {\it The scalar case} $\NU_{0,\sigma}$, where \bei
\item $-2<\sigma < \frac{1}{4}$ for the complementary series;
\item  $\frac{1}{4} \leq \sigma$ for the principal series. \ei
\item {\it The spinorial case} $\NU_{p,\sigma}$, $p>0$, where \bei
\item $0<\sigma < \frac{1}{4}$, $p=1,2, \cdots$, for the
complementary series, \item $\frac{1}{4} \leq \sigma$, $p=1,2,
\cdots$, for the integer spin principal series, \item $\frac{1}{4} <
\sigma$, $p= \frac{1}{2}, \frac{3}{2}, \frac{5}{2} \cdots$,
for the half-integer spin principal series. \ei
\ei
\ei

\setcounter{paragraph}{0}
\subsection{The anti de Sitter group, its unitary irreducible representations, and their physical interpretation}

The anti de Sitter relativity group is $G=SO_{0}(2,3)$, i.e. the component
connected to the identity of the ten-dimensional pseudo-orthogonal
group $SO(2,3)$. Like for dS, a realization of the Lie algebra is that
one generated by the ten Killing vectors
\beq
\label{kilads}
K_{\alpha \beta} = x_{\alpha}\partial_{\beta} -
x_{\beta}\partial_{\alpha}. \eeq
Contrarily to dS, there is one globally time-like Killing
vector in anti de Sitter, namely $K_{50}$. On the other hand, the compact nature of the associated one-parameter group (it is just $SO(2) \simeq U(1)$ or its double covering) can raise problems \cite{AvIsh}. The latter can be circumvented by dealing with the universal covering $\widetilde{G}=\widetilde{SO_{0}(2,3)}$ in  which the ``time'' $SO(2)$ subgroup becomes $\R$.
The two-covering of the anti de Sitter group is the symplectic
$Sp(4, \R)$ group, which  is needed when dealing with half-integer spins.
Here,  the UIR's of
the de anti de Sitter group $Sp(4,\R)$ which are physically meaningful are found in the holomorphic discrete series and in its lower limits. Like in dS, the infinitesimal generators read as:
  \beq
  \label{genrep}
K_{\al \be} \lga L_{\al \be} = M_{\al \be} + S_{\al \be},
  \eeq
where the orbital part is $M_{\al \be}=-i(x_{\al}\partial_{\be} -
x_{\be}\partial_{\al})$ and the spinorial part $S_{\al \be}$ acts on
the indices of functions in a certain permutational way.
 In the case of the discrete series and its lower limit, these UIR's are denoted $D(\varsigma,s)$ with $2s \in \N$ and $ \varsigma \geq s+1$ (at the exception of a few cases). The label   $s$ is for spin (it plays the role of the dS $p$) and $\varsigma$ for lowest ``energy'' (to some extent it plays the role of the dS $q$). For UIR in the strictu senso discrete series of $Sp(4,\R)$, the parameter $\varsigma$ is such that $2\varsigma \in \N$ whilst for ``discrete'' series UIR of the universal covering  $\widetilde{SO_{0}(2,3)}$ this parameter assumes its values in $[s+1, \infty)$. Here too, there are
two Casimir operators, the eigenvalues of which determine completely
the UIR's. With our parameters, they read as
\beq
\label{cas1}
Q^{(1)} = - \frac{1}{2} L_{\al \be}L^{\al \be}, \eeq
with eigenvalues

\begin{equation}
\label{eigenads1}
\langle Q^{(1)} \rangle_{\mathrm{AdS}} =  s(s+1) + \varsigma(\varsigma - 3),
\end{equation}
 and
\begin{equation}
\label{cas2}
Q^{(2)} = - W_{\al}
W^{\al}, \ W_{\al} = - \frac{1}{8}\epsilon_{\al \be \ga \delta \eta}
L^{\be \ga}L^{\delta \eta},
\end{equation}
with eigenvalues
\begin{equation}
\label{eigenads2}
\langle Q^{(2)} \rangle_{\mathrm{AdS}} = -s(s+1)(\varsigma-1)(\varsigma - 2).
\end{equation}

Among the AdS UIR $D(\varsigma,s)$, one must distinguish between those for which $\varsigma > s+1$,   and the following important limit cases
\bei
\item {\it The limit scalar cases} $D(1,0)$ and $D(\frac{1}{2}, 0)$. The latter is called the ``Rac'' \cite{flafron}.

\item {\it
The limit spinorial or tensorial cases}  $D(s+1,s)$ and $D(1, \frac{1}{2})$. The latter is called the ``Di'' \cite{flafron}.\ei

\section{Minkowskian content of dS and AdS elementary systems: contraction results and the Garidi mass}

Now, we wish to go further into the interpretative problem of mass in a dS/AdS background.
The crucial question to be addressed concerns the interpretation of the dS/AdS
UIR's (or quantum AdS and dS elementary systems) from a (asymptotically) minkowskian
point of view. We mean by this the study of the contraction limit
$\vka\to 0$ or equivalently $\Lambda \to 0$ of these representations, which is the quantum counterpart
of the following
geometrical and group contractions.
\paragraph{Flat limit of de Sitter geometry}

\bei
\item $\lim _{ \vka\to 0} M_{dS} = M_0$, the Minkowski spacetime tangent
to $M_{dS}$ at, say, the de Sitter origin point $O_{dS}= (0, \vec{0}, \vka^{-1})$, since then $ M_{dS} \ni x \underset{\vka\to 0}{\approx} (t, \vec{r} = r\, \vec{n}, \vka^{-1})$ from Equation (\ref{coordh}).
\item $\lim _{  \vka\to 0}Sp(2,2) = {\cal P}^{\uparrow}_{+} (1, 3) = M_0
\SD SL(2,\C)$, the Poincar\'e group.
\ei
As a matter of fact, the ten de Sitter Killing vectors (\ref{kil}) contract
to their Poincar\'e counterparts $K_{\mu \nu}$, $\Pi_{\mu}$, $\mu =
0, 1, 2, 3$, after rescaling the four $K_{4\mu} \lga \Pi_{\mu} = \vka
K_{4\mu} $.

\paragraph{Flat limit of anti de Sitter geometry}

\bei
\item $\lim _{ \vka\to 0} M_{AdS} = M_0$, the Minkowski spacetime tangent
to $M_{AdS}$ at, say, the de Sitter origin point $O_{AdS} = (0, \vec{0}, \vka^{-1})$, since then $ M_{AdS} \ni x \underset{\vka\to 0}{\approx} (t = \tau, \vec{r} = r \vec{n}, \vka^{-1})$ from Equation (\ref{adscoordh})
\item $\lim _{  \vka\to 0}Sp(4,\R) = {\cal P}^{\uparrow}_{+} (1, 3) = M_0
\SD SL(2,\C)$.
\ei
Like above, the ten de Sitter Killing vectors (\ref{kilads}) contract
to their Poincar\'e counterparts $K_{\mu \nu}$, $\Pi_{\mu}$, $\mu =
0, 1, 2, 3$, after rescaling the four $K_{5\mu} \lga \Pi_{\mu} = \vka
K_{5\mu} $.
\setcounter{paragraph}{0}
\subsection{Contraction limits de Sitter $\to$ Minkowski}

We have to distinguish between the Poincar\'e massive and
massless cases. We shall denote by ${\cal P}^{\gtrless}(m,s)$
the positive (resp. negative) energy Wigner UIR's of the Poincar\'e
group with mass $m$ and spin $s$. We should here insist on the non-ambiguous definition of minkowskian mass through the mass $m$ label of a UIR of the Poincar\'e group. On the other hand, the notion of mass in ``desitterian Physics'' is  confusing. An  interesting discussion and precisions on  this matter is found in  the  work by Garidi \cite{GAR1},  in which the following  ``mass'' formula has been proposed in terms of the dS RUI parameters $p$ and $q$:
\begin{equation}
\label{garidimass}
m^2_H= \langle Q^{(1)} \rangle_{\mathrm{dS}} - \langle Q^{(1)}_{p=q} \rangle_{\mathrm{dS}}= [(p - q)(p + q - 1)] \hbar^2H^2/c^4.
\end{equation}
This formula is natural in the sense that when the second-order wave equation
\begin{equation}
\label{dswaveeq}
\left(Q^{(1)} - \langle Q^{(1)} \rangle_{\mathrm{dS}}\right)\varphi = 0,
\end{equation}
obeyed by rank $r$ tensor fields carrying a dS UIR, is written in terms of the Laplace-Beltrami operator
$\Box_H$ on the dS manifold, one gets (in units $\hbar = 1 = c$)
\begin{equation}
\label{dswaveeq}
\left(\Box_H  + H^2 r(r+2) + H^2 \langle Q^{(1)} \rangle_{\mathrm{dS}}\right)\varphi = 0.
\end{equation}
Moreover, the minimal value assumed by  the eigenvalues of the first Casimir in the set of RUI in the discrete series is precisely reached at $p=q$, which corresponds to the ``conformal'' massless case, as it will shown below. The Garidi mass has the advantages to encompass all  mass formulas introduced within a de-sitterian context, often in a purely mimetic way in regard with their minkowskian counterparts.

Whenever $ \langle Q^{(1)} \rangle$ does not correspond to a UIR with unambiguous minkowskian interpretation, one can still use $m^2_H$ but without referring to a minkowskian meaning.

For the Poincar\'e massless case we shall make use of similar
notation
${\cal P}^{\gtrless}(0,s)$
where $s$ reads for helicity. In the latter case, conformal
invariance leads us to deal also with the discrete series
representations
(and their lower limits) of the (universal covering of the)
conformal group or its double covering $SO_0(2,4)$ or its fourth
covering $SU(2,2)$. These UIR's are denoted in the sequel by
${\cal C}^{\gtrless}(\varsigma,j_1, j_2)$, where $(j_1,j_2) \in
\N/2 \times \N/2$ labels the UIR's of the $SU(2) \times SU(2)$ subgroup and
$\varsigma$ stems for the positive (resp. negative) conformal energy. The
de Sitter contraction limits are summarized in 
diagrams below.
\paragraph{dS massive case}
Solely the principal series representations  are
involved here (from where the name of de Sitter ``massive
representations''). Introducing the representation  parameter $\nu \in \R $ through $q = \frac{1}{2} + i \nu$ or equivalently $\sigma
= \nu^2 + 1/4$ (note that dS UIR corresponding to $\nu$ and $-\nu$ are equivalent) and for a spin $s$, the Casimir eigenvalue and Garidi mass read respectively:
\begin{equation}
\label{PSdSCas}
\langle Q^{(1)} \rangle_{\mathrm{dS}} = -s(s+1) + \nu^2 + \frac{9}{4},
\end{equation}
\begin{equation}
\label{GarmassdS}
\ m_H= \frac{\hbar H}{c^2} \,\sqrt{(s-\frac{1}{2})^2 + \nu^2}.
\end{equation}

Let $m$ be a  mass in the Poincar\'e-Minkowski sense defined by
\begin{equation}
\label{dsnumass1}
m= \vert \nu \vert  \hbar H/c^2 = \vert \nu \vert \vka\frac{\hbar}{c} = \vert  \nu \vert  \frac{\hbar}{c} \sqrt{\frac{ \vert \Lambda\vert}{3}}.
\end{equation}
Then   we have the following general result on contraction of dS principal series representations:
\cite{[MINI],GAHURE}
\beq
\label{dscontr1}
{\NU}_{s,\sigma}\underset{\vert \nu \vert  \vka= \frac{mc}{\hbar} }{\underset{\vka\to 0, \vert \nu \vert\to \infty}{\lga}} {c_>\cal P}^{>}(m,s)
\oplus c_<{\cal P}^{<}(m,s),
\eeq
where one of the ``coefficients'' among $c_<, \, c_>$ can be fixed to 1 whilst the other one will vanish. Note that $m = m_H + O(\vka)$.
Note also here the evidence of the energy ambiguity in de Sitter relativity,
exemplified by the possible breaking of dS irreducibility into a direct sum of
two Poincar\'e UIR's with positive and negative energy respectively.
This phenomenon is linked to the existence in the de Sitter group of
a specific discrete symmetry, precisely $\ga_0 \in Sp(2,2)$, which
sends any point $(x^0, {\cal P}) \in M_{dS}$ (with the notations of
(2.7)) into its mirror image $(x^0, -{\cal P}) \in M_{dS}$ with respect
to the $x^0$-axis. Under such a symmetry the four generators
$L_{a0}$, $a = 1,2,3,4$, (and particularly $L_{40}$ which contracts
to energy operator!) transform into their respective opposite
$-L_{a0}$, whereas the six $L_{a b}$'s remain unchanged.

\paragraph{dS massless (conformal) case}

Here we have $m_H = 0$ for all involved representations.
Now, we must distinguish between the
scalar massless case, which involves the unique complementary series
UIR $\NU_{0,0}$ (for which $\langle Q^{(1)} \rangle_{\mathrm{dS}} = 2$) to be contractively Poincar\'e significant, and the
spinorial case where are involved all representations
$\Pi^{\pm}_{s,s}, \ s>0$ for which $\langle Q^{(1)} \rangle_{\mathrm{dS}} = -2(s^2 -1)$ and lying at the lower limit of the discrete
series. The arrows $\hookrightarrow $ below designate unique
extension.

\subparagraph{dS scalar massless case }
{\footnotesize
\beq \left. \begin{array}{ccccccc}
&   & {\cal C}^{>}(1,0,0)
& &{\cal C}^{>}(1,0,0) &\hookleftarrow &{\cal P}^{>}(0,0)\\
\NU_{0,0} &\hookrightarrow & \oplus
&\stackrel{\vka= 0}{\longrightarrow} & \oplus & &\oplus \\
&   & {\cal C}^{<}(-1,0,0)&
& {\cal C}^{<}(-1,0,0) &\hookleftarrow &{\cal P}^{<}(0,0),\\
\end{array} \right. \eeq
}
\subparagraph{dS spinorial massless case }
{\footnotesize
\beq \left. \begin{array}{ccccccc}
&   & {\cal C}^{>}(s+1,s,0)
& &{\cal C}^{>}(s+1,s,0) &\hookleftarrow &{\cal P}^{>}(0,s)\\
\Pi^+_{s,s} &\hookrightarrow & \oplus
&\stackrel{\vka= 0}{\longrightarrow} & \oplus & &\oplus \\
&   & {\cal C}^{<}(-s-1,s,0)&
& {\cal C}^{<}(-s-1,s,0) &\hookleftarrow &{\cal P}^{<}(0,s),\\
\end{array} \right. \eeq
\beq \left. \begin{array}{ccccccc}
&   & {\cal C}^{>}(s+1,0,s)
& &{\cal C}^{>}(s+1,0,s) &\hookleftarrow &{\cal P}^{>}(0,-s)\\
\Pi^-_{s,s} &\hookrightarrow & \oplus
&\stackrel{\vka= 0}{\longrightarrow} & \oplus & &\oplus \\
&   & {\cal C}^{<}(-s-1,0,s)&
& {\cal C}^{<}(-s-1,0,s) &\hookleftarrow &{\cal P}^{<}(0,-s),\\
\end{array} \right. \eeq
}

Finally, all
other representations have either non-physical Poincar\'e contraction
limit or have no contraction limit at all. In  particular, we have for the so-called
\emph{massless minimally coupled field} which corresponds to the UIR
$\Pi^+_{1,0}$ lying at the lowest limit of the discrete series the following values for
Casimir eigenvalue and Garidi mass:
\begin{equation}
\label{mmcvalue}
\langle Q^{(1)} \rangle_{\mathrm{dS}} = 0, \ m_H = 0.
\end{equation}

This representation, and hence the corresponding field, is exceptional under many aspects. First, it is the only one among all non  massless dS representations for which the Garidi mass vanishes, and it is part of an indecomposable structure issued from the existence of (constant) gauge solutions to (\ref{mmcvalue}).   Secondly,  it 
has been playing a crucial role in inflation theories, it 
is part of the Gupta-Bleuler structure  for the massless spin 1 dS field (de Sitter QED) described by the UIR's $\Pi^+_{1,1}$ \cite{gagarouta}, and it is the elementary brick for the construction of the massless spin 2 dS fields (de Sitter linear gravity) described by the UIR's $\Pi^+_{2,2}$ \cite{garouta}. Finally, the corresponding covariant quantum field theory requires a specific treatment  due precisely to  its indecomposable nature \cite{gareta}.

We have now reached the point at which one can reinterpret the mass content of Eqs. (\ref{4.10},\ref{gravmass1}) in the above framework. Indeed, comparing Eqs. (\ref{4.10}) and  (\ref{dswaveeq}) it has been proved   in \cite{GAR1} that the following relation holds true
for spin-2   particles which are massless in the dS sense:
\begin{equation}
\label{garnov}
m^2_g + \frac{2}{3}\Lambda = m^2_H = 0.
\end{equation}

This equation which relates the minkowskian mass to the de Sitter mass $m_H$ gives a precise meaning to the equations (\ref{gravmass},\ref{gravmass1}).

\setcounter{paragraph}{0}
\subsection{Contraction limits anti de Sitter $\to$ Minkowski}


A  ``mass'' formula analogous to the Garidi one can be proposed here in the case of the AdS discrete series. It will give a zero-mass for massless AdS fields:
\begin{align}
\label{adsgaridimass}
\nonumber m^2_H=& \langle Q^{(1)} \rangle_{\mathrm{AdS}} - \langle Q^{(1)}_{\varsigma=s + 1} \rangle_{\mathrm{AdS}}, \\
\mathrm{i.e.}\ m_H&=  \frac{\hbar H}{c^2} \,\sqrt{\left(\vsi - \frac{3}{2}\right)^2 - \left(s - \frac{1}{2}\right)^2}.
\end{align}

With the same notations as above,  the
anti de Sitter contraction limits can be summarized in the following
diagrams.

\paragraph{AdS massive case}
Solely the (holomorphic) discrete series representations  $D(\varsigma,s)$ with $\varsigma > s+1$ are
involved here. Introducing the following relation between  the representation  parameter $\varsigma > s+1$  and a Poincar\'e-Minkowski mass:
\begin{equation}
\label{adsnumass1}
m=  \varsigma   \vka\frac{\hbar}{c} = \varsigma  \frac{\hbar}{c} \sqrt{\frac{ \vert \Lambda\vert}{3}},
\end{equation}
  we have
\cite{adscont}
\beq
\label{adscontr1}
D(\varsigma,s) \underset{\vsi  \vka= \frac{mc}{\hbar} }{\underset{\vka\to 0, \varsigma \to \infty}{\lga}} {\cal P}^{>}(m,s).
\eeq

Note here that there is no  energy ambiguity in anti de Sitter relativity (there are other ambiguities!). If we wished to get the negative energy Poincar\'e representations, we  would instead have chosen the representations in the \emph{antiholomorphic} discrete series (in which the  spectrum of the compact generator $L_{50}$ is bounded above by $-\varsigma, \varsigma > 0$):

\beq\underline{\underline{}} \label{adscontr2} D(-\varsigma,s)
\underset{\vka\to 0, \varsigma \to \infty}{\lga} {\cal P}^{<}(m,s).
\eeq

\paragraph{AdS massless (conformal) case}

Here we must distinguish between the
scalar massless case, which involves the
UIR $D(1,0)$  and the
spinorial-tensorial case in which are involved all representations
$D(s+1,s), \ s>0$ lying at the lower limit of the holomorphic discrete
series. Here, there is no ambiguity concerning energy, but there is ambiguity concerning helicity, since the later  is not defined in AdS. As above, the arrows $\hookrightarrow $ below designate unique
extension.

\subparagraph{AdS scalar massless case}
{\footnotesize
\beq \left. \begin{array}{ccccccc}
D(1,0)& \hookrightarrow & {\cal C}^{>}(1,0,0)
&\stackrel{\vka= 0}{\longrightarrow}  &{\cal C}^{>}(1,0,0) &\hookleftarrow &{\cal P}^{>}(0,0).\\
\end{array} \right. \eeq
}
\subparagraph{AdS spinorial tensorial massless case}
{\footnotesize
\beq \left. \begin{array}{ccccccc}
&   & {\cal C}^{>}(s+1,s,0)
& &{\cal C}^{>}(s+1,s,0) &\hookleftarrow &{\cal P}^{>}(0,s)\\
D(s+1,s) &\hookrightarrow & \oplus
&\stackrel{\vka= 0}{\longrightarrow} & \oplus & &\oplus \\
&   & {\cal C}^{>}(s+1,0,s)&
& {\cal C}^{>}(s+1,0,s) &\hookleftarrow &{\cal P}^{>}(0,-s).\\
\end{array} \right. \eeq
}

Finally, all
other representations have either non-physical Poincar\'e contraction
limit or have no contraction limit at all.
In  particular, we have for the
\emph{Rac} and \emph{Di} fields the following respective values for
Casimir eigenvalue and Garidi mass:
\begin{equation}
\label{Racvalue}
\langle Q^{(1)} \rangle_{\mathrm{AdS}} = -\frac{5}{4}, \ m_H = \frac{\sqrt{3}}{2} \frac{ \hbar H}{c^2}, \ \mbox{(Rac)},
\end{equation}
\begin{equation}
\label{Divalue}
\langle Q^{(1)} \rangle_{\mathrm{AdS}} = -\frac{5}{4}, \ m_H = \frac{ \hbar H}{2c^2}, \ \mbox{(Di)}.
\end{equation}
It should be also noted that, like for de Sitter,  there exists a unique UIR, among all non  massless AdS representations, for which $m_H$ vanishes, namely the UIR $D(2,0)$ in the discrete series.
\setcounter{paragraph}{0}
\section{Non-relativistic contraction limits with track of the curvature: Newton space-times}
dS and AdS  theories involve universal length  ($\sim1/\sqrt{\vert \Lambda \vert}$) and  universal speed $c$. The analysis of their physical content, specially from the point of view of the question of mass, would not be complete if their respective ``non-relativistic'' limit were not considered.
Precisely, other possibilities of contractions from dS and AdS systems exist. There have been listed  by Bacry and L\'evy-Leblond \cite{bajmll} on the basis of symmetry principles and mild physical assumptions, namely, isotropy of  space,
 parity and time-reversal are automorphisms of the 
kinematical group, and 
inertial transformations in any given direction form a non-compact 
group. Hence, besides their common minkowskian-Poincar\'e limit,  the de Sitter and anti de Sitter groups are also  respective deformations of the so-called Newton groups, kinematical groups of the so-called Newton space-times. More precisely, let us perform on the dS (\ref{coordh}) and AdS (\ref{adshyp}) global coordinates  the following ``non-relativistic approximations'' after introducing the ``universal time'' for dS

\begin{equation}
\label{univtime}
\tau = H^{-1} = \frac{1}{\vka c},
\end{equation}

and the universal frequency for AdS,

\begin{equation}
\label{univfreq}
\omega = H = \vka c.
\end{equation}

Note from \cite{pada} the current estimate for the Hubble length: $c/H_0\equiv \vka_0^{-1} \approx 1.2\times 10^{26}$m,
which gives for the Hubble time or estimated age of universe the value $H_0^{-1} \approx 0.4\times 10^{18}\,\mathrm{s}\, \approx 1.27 \times 10^{10}$y.

\paragraph{Newton space-time $\mathcal{N}_+$ as a non-relativistic approximation of the dS space-time}

\begin{figure}[h]
\begin{center}
\includegraphics[width=9cm]{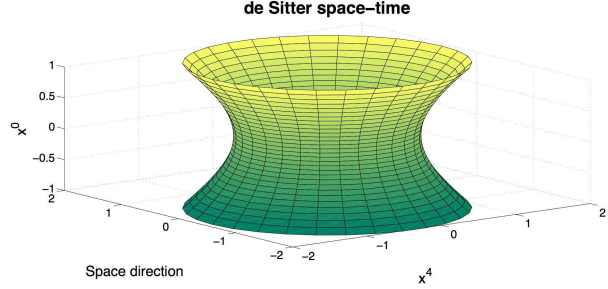}
\caption{ de Sitter space-time as a hyperboloid embedded in $\R^5$}
\label{dS3}
\end{center}
\end{figure}

In terms of the universal time $\tau$ defined in (\ref{univtime})  the dS global coordinates (\ref{coordh}) read as:

\begin{equation}
\label{univtimeds}
x = \left\lbrace \begin{array}{ll}
 x^0       & =  c\,\tau \sinh{(t/\tau)} \\
 \vec{x}      & =   c\,\tau \cosh{(t/\tau)}\sin{\left(r/(c\tau)\right)}\vec{n}\\
 x^4      &  = c\,\tau \cosh{(t/\tau)} \cos{\left(r/(c\tau)\right)}
\end{array} \right.
\end{equation}
The de Sitter space-time is shown in Figure \ref{dS3}.

Its non-relativistic approximation at large $c$ \underline{and} for distances $r \ll c \tau$ is then given in terms of the universal time $\tau$ and the corresponding universal length $R =: c \tau = H^{-1}$ by:
\begin{equation}
\label{nrds}
x = \left\lbrace \begin{array}{ll}
 x^0       &=  R \sinh{(t/\tau)} \\
 \vec{x}     &=   \vec{r} = r\vec{n}\\
 x^4       &=   R \cosh{(t/\tau)}
\end{array} \right.
\end{equation}

\begin{figure}[h]
\begin{center}
\includegraphics[width=9cm]{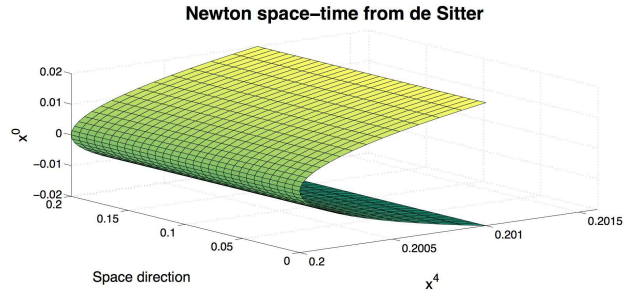}
\caption{Newton $\mathcal{N}_+$ as a nonrelativistic approximation of the de Sitter space-time}
\label{Newton1}
\end{center}
\end{figure}

This describes (see Figure \ref{Newton1}) a kind of hyperbolic cylindrical sheet in $\R^5$ opening in the direction of the positive values of $x^4$.

Of course we recover the galilean space-time at times $t \ll \tau$.

Note that for distances of the order of the universal length $r \approx \pi c\, \tau = \pi R$ one would get negative values $x^4 = - c\, \tau \cosh{(t/\tau)}$  which means that the reached Newton space-time is a mirror hyperbolic cylindrical sheet under $x^4 \to -x^4$.

\paragraph{Newton space-time $\mathcal{N}_-$ as a non-relativistic approximation of the AdS spacetime}

In terms of the universal frequency $\omega$ defined in (\ref{univfreq})  the AdS global coordinates (\ref{adshyp}) read as:

\begin{equation}
\label{univfreqads}
x = \left\lbrace \begin{array}{rl}
 x^0       &=  c\,\omega^{-1} \sin{(\omega t)} \cosh{\left(\omega r/c \right)} \\
 \vec{x}     & =  c\,\omega^{-1}\sinh{\left(\omega r/c\right)}\vec{n}\\
 x^5       & =  c\,\omega^{-1} \cos{(\omega t)} \cosh{\left(\omega r/c\right)}
\end{array} \right.
\end{equation}

\begin{figure}[h]
\begin{center}
\includegraphics[width=9cm]{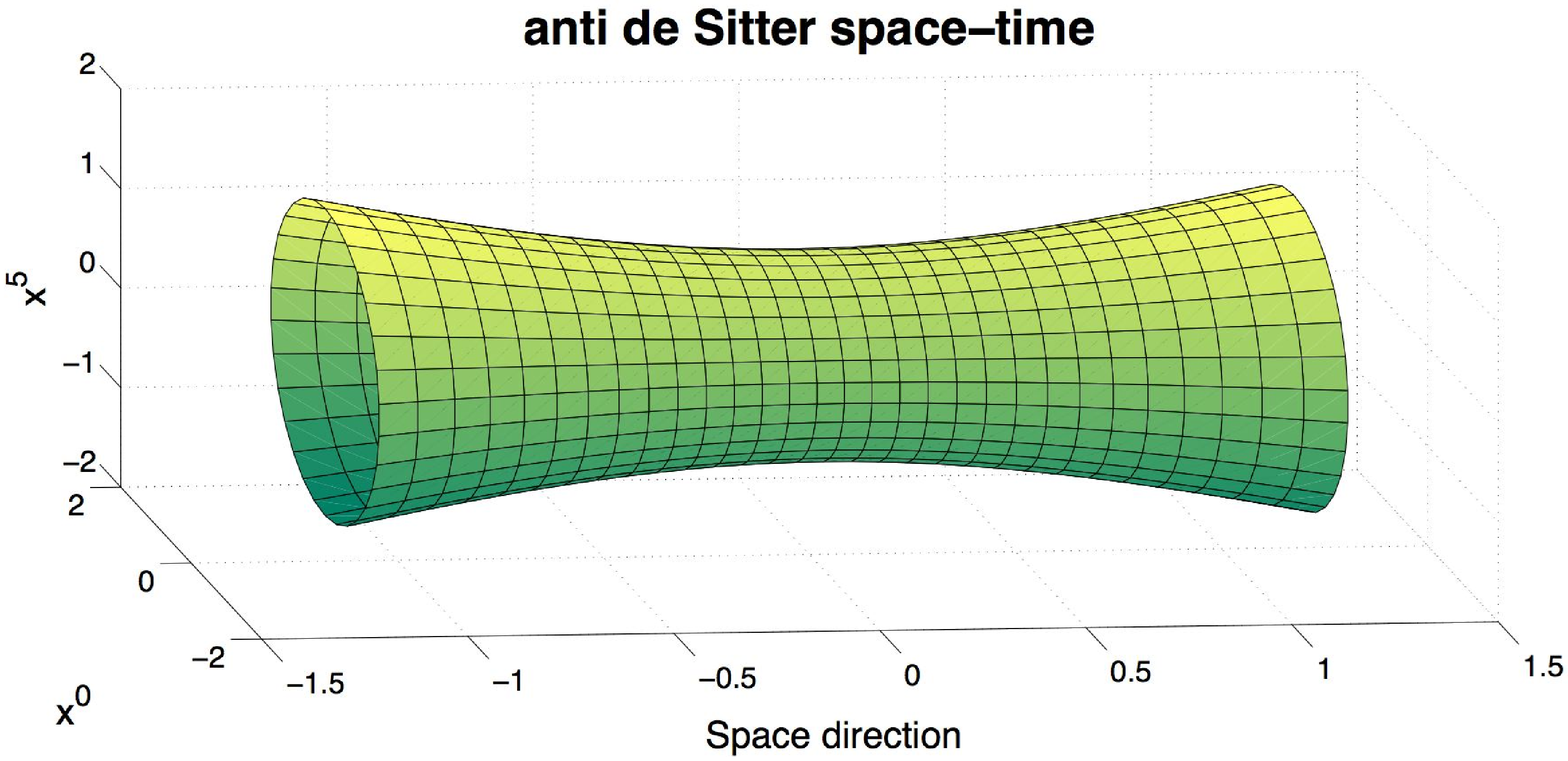}
\caption{ anti de Sitter space-time as a hyperboloid embedded in $\R^5$}
\label{AdS1}
\end{center}
\end{figure}

The anti de Sitter space-time is shown in Figure \ref{AdS1}.

Its non-relativistic approximation at large $c$ \underline{and} for distances $r \ll c\, \omega^{-1}$ is then given in terms of the universal frequency $\omega$ and the corresponding universal length $R =: c \,\omega^{-1} = H^{-1}$ by:
\begin{equation}
\label{nrads}
x = \left\lbrace \begin{array}{ll}
 x^0       &=  R \sin{(\omega t)} \\
 \vec{x}      & =  \vec{r} = r\vec{n}\\
 x^5      & =  R \cos{(\omega t)}
\end{array} \right.
\end{equation}

\begin{figure}[h]
\begin{center}
\includegraphics[width=9cm]{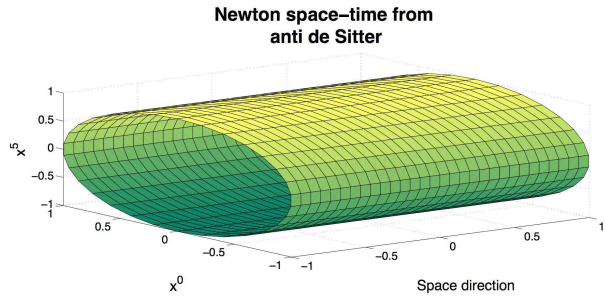}
\caption{Newton $\mathcal{N}_-$ as a nonrelativistic approximation of the anti de Sitter space-time}
\label{Newtads}
\end{center}
\end{figure}

This describes (see Figure  \ref{Newtads})) a   cylindrical hypersurface in $\R^5$ with axis in the direction of spatial  $\vec{x}$.

Again we recover the galilean space-time at times $t \ll \omega^{-1}$.

\setcounter{paragraph}{0}
\section{Contraction limits and the question of interpretation of mass in dS and AdS  space-times}

Actually, both contraction formulas (\ref{dsnumass1}) and (\ref{adsnumass1}), established on a group irrep. level,  are by far restrictive. Of course, they give abstract dimensionless parameters $\nu$ and $\vsi$ labelling respectively the UIR's of the dS and AdS groups a status of physical quantity in terms of measurable other physical quantities, like a mass $m$ and a cosmological constant $\Lambda$ (universal?), and of universal constants, like $c$ and $\hbar$. However, given a minkowskian mass $m$ and a ``universal'' length $R = \vka^{-1}=:\sqrt{3/\vert \Lambda\vert}= c\,H^{-1}=c\tau$, nothing prevents us to consider those two quantities, specific of a ``physics'' in constant-curvature space-time, as meromorphic functions of the dimensionless physical \footnote{in the minkowskian sense!} quantity, expressed in terms of  various other quantities introduced in this paper,
\begin{align}
\label{nodimm}
\nonumber \vth \equiv  \vth_m \stackrel{\mathrm{def}}{=} &\frac{\hbar \vka}{mc} = \frac{\hbar}{Rmc}
= \frac{\hbar \sqrt{\vert \Lambda} \vert}{\sqrt{3}mc} \\ = &\frac{\hbar\, H}{ mc^2}= \frac{\hbar}{\tau mc^2} = \frac{\hbar \omega}{mc^2} = \frac{m_{\Lambda}}{\sqrt{3}m}.
\end{align}
Note that this quantity is also the ratio of the Compton length of the  minkowskian object of mass $m$ considered at the limit with the universal  length $R = \vka^{-1}$ yielded by dS or AdS geometry. It  reduces to   $\vka/m$ in units $\hbar = 1 = c$. It is just equal to $1$ for  $m = m_{\Lambda}/\sqrt{3}$ where $m_{\Lambda}$ was introduced in (\ref{4MAr3}) apropos  of the two mass scales. We give in Table \ref{tab:table1}  the values assumed by the quantity $\vth$ when $m$ is taken as  some known masses and $\Lambda$ (or $H_0$) is given its present day estimated value. We  easily  understand from this table that the currently estimated value of the cosmological constant has no practical effect on our familiar massive fermion or boson fields. Contrariwise, adopting  the  de Sitter point of view appears as inescapable when we deal with  infinitely small masses, as is done in standard inflation scenario. 

\begin{table}[h]
\caption{Estimated values of the dimensionless physical quantity $\vth \equiv \vth_m =:  \frac{\hbar \sqrt{\vert \Lambda} \vert}{\sqrt{3}mc}=  \frac{\hbar\, H}{ mc^2} \approx 0.293\times 10^{-68}\times m_{\mathrm{kg}}^{-1} $ for some known masses $m$ and the present day estimated value of the Hubble length $ c/H_0 \approx 1.2\times 10^{26} \mathrm{m}$ \cite{pada}.}
 { \begin{tabular}[c]{|c|c|}\hline
Mass $m$ & $\vth_m \approx $\\ \hline
$m_{\Lambda}/\sqrt{3}\approx 0.293\times 10^{-68}$kg & 1 \\ \hline
up. lim. photon mass $m_{\gamma}$& $0.29 \times 10^{-16}$\\ \hline
up. lim. neutrino mass $m_{\nu}$& $0.165 \times 10^{-32}$\\ \hline
electron mass $m_e$& $0.3 \times 10^{-37}$\\ \hline
proton mass $m_p$& $0.17  \times 10^{-41}$ \\ \hline
$W^{\pm}$  boson mass & $0.2  \times 10^{-43}$\\ \hline
Planck mass $M_{Pl}$& $0.135  \times 10^{-60}$\\ \hline
\end{tabular}}
\label{tab:table1}
\end{table}

Now, we may consider the following Laurent expansions of $\nu$ (for the dS prinicipal series)  and $\vsi$ (for the AdS discrete series) in a certain neighborhood of $\vth = 0$:
\begin{align}
\label{laurentnu}
 \nu =    & \nu (\vth)=   \frac{1}{\vth} + e_0 + e_1 \vth + \cdots e_n \vth^n + \cdots\\
 \label{laurentsi}  \vsi =    & \vsi (\vth)=  \frac{1}{\vth} + f_0 + f_1 \vth + \cdots f_n \vth^n + \cdots , \\
\nonumber \vth \in & (0, \vth_1) \ \mbox{convergence interval},
\end{align}
where the $e_n, f_n$ are pure number to be determined. We should be aware that nothing is changed in the  contraction formulas (\ref{dscontr1}) and (\ref{adscontr1}) from the point  of view of a minkowskian observer, except that we allow to consider  positive as well as negative values of $\nu$ in a (positive) neighborhood of $\vth = 0$: multiply (\ref{laurentnu}) and (\ref{laurentsi}) by $\vth$ and go to the limit $\vth \to 0$. We recover asymptotically the relations (\ref{dsnumass1}) and (\ref{adsnumass1}).

As a matter of fact, the Garidi mass (\ref{GarmassdS}) in the dS case or the mass formula (\ref{adsgaridimass}) proposed for the AdS case are perfect examples of such expansions since they can be rewritten as the following expansions in the parameter $\vth \in ( 0,1/\vert s- 1/2 \vert]$:
\begin{align}
   \nonumber \nu & =  \sqrt{\frac{1}{\vth^2} - (s-1/2)^2}\\
\label{dsnu}   &= \frac{1}{\vth} -(s-1/2)^2\left(\frac{\vth}{2} + O(\vth^2)\right),\\  
\nonumber \vsi & =  \frac{3}{2} +  \sqrt{\frac{1}{\vth^2} + (s-1/2)^2} \\
\label{dszeta} &= \frac{1}{\vth} + \frac{3}{2} + (s-1/2)^2\left(\frac{\vth}{2}  + O(\vth^2)\right).
\end{align}
Note the particular symmetric place occupied by the spin $1/2$ case with regard to the scalar case $s=0$ and the boson case $s=1$.

Hence, we can  tell something  more on the number $f_0$ introduced for the anti de Sitter case, and this represents one more motivation for exploring further the possibilities offered by the above expansions.  From the preceding section,
 an AdS  scalar elementary system can be viewed as a deformation of
both a  relativistic free particle with rest energy $mc^2$ and a harmonic oscillator with rest energy
$\frac{3}{2}\hbar \omega$. It has thus been proven in \cite{gare1} in the $1+1$-dimensional case  the following
\begin{equation}
\vsi=\frac{mc}{\hbar \vka}+\frac{1}{2}+O(\vka),\label{choixeta}
\end{equation}
which means precisely that $ f_0 = 1/2$ in this case. From which is derived the following expansion of the energy of a scalar ``massive'' AdS elementary system from a minkowskian tangent point of view:
\begin{equation}
\label{adsenergy}
E_{AdS} = mc^2 + \frac{1}{2}\hbar \omega + O(\vka).
\end{equation}
The extension of the proof to the $3+1$-dimensional case is  straightforward and the result is in perfect agreement with the content of the expansion (\ref{dszeta}) concerning the appearance of the constant term $3/2$:
\begin{equation}
E_{AdS} = mc^2 + \frac{3}{2}\hbar \omega + O(\vka),\label{choixeta3+1}
\end{equation}

On the other hand, the situation of dS relativity with regard to its both Poincar\'e and Newton limits is less tractable. It is well exemplified by the absence of any constant term in (\ref{dsnu}).

Let us insist once more on  the very peculiar position occupied by the spin $s = 1/2$   since then we exactly have from   (\ref{dsnu}-\ref{dszeta}):
\begin{equation}
\label{sdemi}
\nu =  \frac{1}{\vth} \ \mbox{and} \  E_{AdS} = mc^2 + \frac{3}{2}\hbar \omega.
\end{equation}
So, in this particular case, the range of possible values for $\vartheta$ is the positive real axis $(0, \infty)$.

We give in Appendix another example of such expansions in order to illustrate the fact that contraction procedures are far from being bi-univocal as well  physically  as  mathematically.

\section{Conclusion}

In this paper we have presented arguments in favor of giving the
bare $\Lambda$ the status of a fundamental constant, as much small
it can be. A basic physical quantity like  mass  has
then to be reexamined from  a new physical point of view. We face two possibilities: either one starts from a minkowskian background, i.e. adopting the  EGR-1 point of view, and turn on gravity (in particular in order to get the (anti) de Sitter structure) or one starts directly from within the framework of a (anti) de Sitter geometry (EGR-2 point of view). In the first case, an arbitrary field (say, for the case of spin-2) has a well-defined mass due to Poincar\'e invariance. Coupling this field to gravity in a (anti) de Sitter background makes the mass of the field acquire an extra contribution. 

However, there is a very particular case in which the expression of the total mass (the original minkowskian one plus an extra contribution of the  (anti) de Sitter structure) implies that the field has only two degrees of freedom, as one should expect for the case of a massless tensor field. This phenomenon  happens precisely when one performs a small perturbation of the EGR-2 equations of motion. One is thus led to associate such a special case to the mass of the graviton. Nevertheless, we insist on the fact that the graviton is not a particle that is massless in the EGR-1 (i.e. minkowskian) sense. The graviton is instead a massless spin-2 particle in the sense of (anti-) de Sitter relativity. Let us also stress on the fact that this interpretation, through the equation 
\begin{equation}
\label{gravmass2}
m^2_H = m^2_g + \frac{2}{3} \Lambda = 0,
\end{equation}
does not allow any acausal propagation. The latter would  be a misinterpretation  issued from the EGR-1 point of view in the $\Lambda > 0$ case. 

The current observation of an accelerated universe points in favor of a desitterian arena. In consistency with this fact, we propose to reexamine carefully the equation (\ref{gravmass2}) which makes the bridge between the bare cosmological term 
$\Lambda$ and a ``mass'' attributed to the graviton.

\section*{Acknowledgement}

This work was partially supported by the Brazilian agency Conselho
Nacional de Desenvolvimento Cient\'{\i}fico e Tecnol\'ogico
(CNPq). M. N.   acknowledges the Laboratory APC of the University Paris 7-Denis Diderot  and  J-P. G.  acknowledges the CBPF  for financial supports.

\appendix
\section{A pedagogical example of contraction expansion}

\begin{figure}[h]
\begin{center}
\includegraphics[width=9cm]{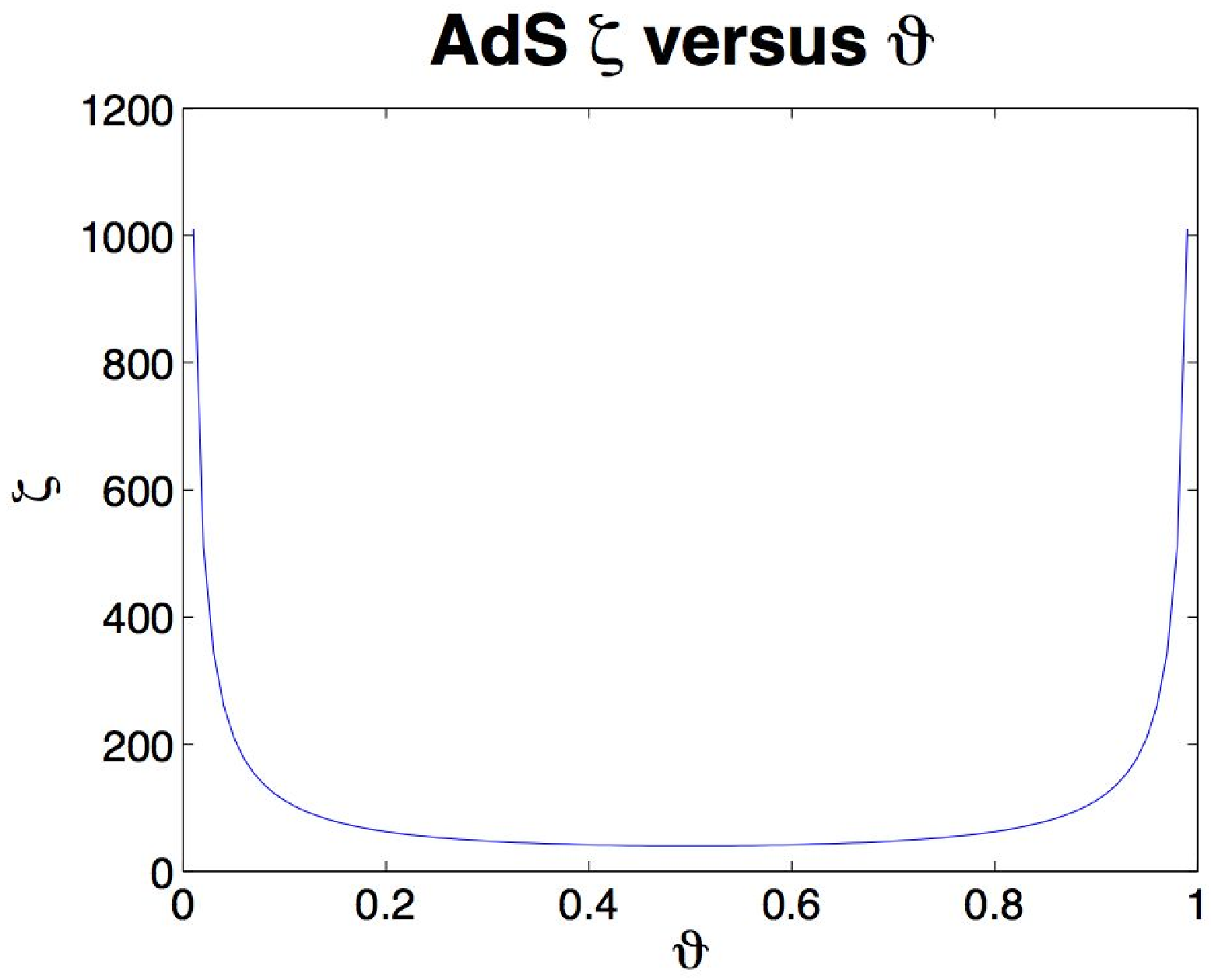}
\caption{An example (with $\alpha = -1$ and $\vth_1 = 1$) of variation of the AdS UIR parameter $\varsigma$ in terms of the $\vartheta$ variable.}
\label{AdScont}
\end{center}
\end{figure}

We here consider a speculative  example in which we suppose that the expansions (\ref{laurentnu}-\ref{laurentsi}) are actually those in the interval $\vth \in (0,\vth_1)$ for the specific function

\begin{equation}
\label{specexp}
\nu \ \mbox{or} \ \vsi = \frac{1}{\vth} + \frac{\alpha}{\vth - \vth_1}, \ \mbox{i.e.} \
e_n = -\frac{\alpha}{\vth_1^{n+1}} = f_n. 
\end{equation}
Two cases have to be considered:
\paragraph{Case $\alpha \leq 0$:}
Here, the function (\ref{specexp}) is positive and assumes the  minimal value $(1+\sqrt{\vert \alpha\vert})^2/\vth_1$ at $\vth = \vth_1/(1 + \sqrt{\vert \alpha\vert})$ (Figure \ref{AdScont}). This case is most appropriate to the AdS case in which the UIR parameter $\vsi$ is bounded below by the unitarity limits $\vth_m = s+1$ for spin $s\geq 1$, $\vth_m = 1$ for spin $1/2$ (``Di''), and $\vth_m = \frac{1}{2}$ (``Rac'') for spin $s = 0$.

\begin{figure}[h]
\begin{center}
\includegraphics[width=9cm]{dScontraction.eps}
\caption{An example (with $\alpha = 1$ and $\vth_1 = 1$) of variation of the dS UIR parameter $\nu$ in terms of the $\vartheta$ variable.}
\label{dScont}
\end{center}
\end{figure}

\paragraph{Case $\alpha > 0$:}
Here, the function (\ref{specexp}) has no minimum and changes its sign at the value $\vth = \vth_1/(1 + \alpha)$ (Figure \ref{dScont}). This case is appropriate to dS only in which the UIR parameter $\nu$ can assume all real values, knowing that for opposite values of it we get equivalent dS UIR.
Now let us examine in both cases what happens at the right-hand limit $\vth \to \vth_1-0$ of the interval. It can be viewed also as a zero-curvature limit, since we have
\begin{align*}
\nonumber \lim_{\vth \to \vth_1-0}(\vth - \vth_1)& (\nu \ \mbox{or} \ \vsi) = \\
 \lim_{\vka\to 0}&\frac{\hbar \vka}{c}\left( \frac{1}{m}- \frac{1}{m_1}\right)(\nu \ \mbox{or} \ \vsi) = \alpha,
\end{align*}
where $m_1$ is a mass corresponding to the limit $\vsi_1= \dfrac{\hbar \vka}{m_1 c}$
This means that we obtain at the contraction limit a minkowskian elementary system with
``corrected'' mass $mm_1/(m_1 - m)$.

\end{document}